# Spectro-microscopy of single and multi-layer graphene supported by a weakly interacting substrate

Kevin R. Knox[1,2], Shancai Wang[2], Alberto Morgante[3,4] Dean Cvetko[3,5], Andrea Locatelli[6], Tevfik Onur Mentes[6], Miguel Angel Niño [6], Philip Kim[1], R. M. Osgood Jr. [2]*

[1] *Department of Physics, Columbia University, New York, New York 10027, USA*

[2] *Department of Applied Physics, Columbia University, New York, New York 10027, USA*

[3] *Laboratorio TASC-INFM 34012, Basovizza, Trieste, Italy*

[4] *Department of Physics, Trieste University, Trieste, Italy*

[5] *Faculty for Mathematics and Physics, University of Ljubljana, Ljubljana, Slovenia.*

[6] *Elettra - Sincrotrone Trieste S.C.p.A., 34012 Basovizza, Trieste, Italy*

**email: osgood@columbia.edu*

**Abstract**

We report measurements of the electronic structure and surface morphology of exfoliated graphene on an insulating substrate using angle-resolved photoemission and low energy electron diffraction. Our results show that although exfoliated graphene is microscopically corrugated, the valence band retains a massless fermionic dispersion, with a Fermi velocity of ~$10^6$ m/s. We observe a close relationship between the morphology and electronic structure, which suggests that controlling the interaction between graphene and the supporting substrate is essential for graphene device applications.

Since its recent experimental realization [1], graphene has been the subject of intense research interest, primarily because of its unusual electronic transport properties. Its unique conical valence and conduction bands—termed Dirac cones—mimic the dispersion of relativistic massless fermions [2]. This leads to intriguing new transport phenomena—for example, a half integer quantum hall effect and Berry's phase have recently been discovered in micro-mechanically exfoliated graphene[3,4]. However, due to limitations in the size of exfoliated graphene flakes, previous experimental investigations of the valence band of graphene have been carried out on layers grown epitaxially on SiC [5,6,7]. Although epitaxial graphene shares many of the same physical properties as exfoliated graphene, it is not a true self-supporting 2D crystal, but is intrinsically bound to the substrate. Interaction with the SiC substrate modifies the graphene lattice in this system [8,9,10], which may have an effect on the measured electronic properties. Additionally, graphene samples grown on SiC are not uniform; rather, they consist of disconnected domains several hundred nanometers in size, each differing in thickness [8,9,10]. On the other hand, graphene sheets prepared by micro-mechanical cleavage of bulk graphite are high quality, uniform in thickness and are true 2D crystals, capable of maintaining crystalline order without a supporting substrate. Exfoliated graphene on $SiO_2$ is the system of choice for the majority of transport experiments as it is relatively easy to gate and has shown the most interesting and impressive electrical transport properties. When supported on $SiO_2$, exfoliated graphene interacts minimally with the substrate; however, given the small size of currently available samples, microscopic probing techniques are required to study these sheets. Here, we report measurements of the surface morphology and electronic structure of exfoliated graphene on a weakly interacting substrate using low energy electron diffraction and angle-resolved photoemission performed on a microscopic scale. Our diffraction measurements show that monolayer and multilayer graphene sheets are not atomically flat, but microscopically corrugated. Our photoemission measurements confirm that the electronic structure is well described by the one-orbital tight binding model—dispersion is linear in the vicinity of the Dirac point and our measured Fermi velocity is very close to the theoretically predicted value of $10^6$ m/s.

Graphene samples were extracted by micro-mechanical cleavage from Kish graphite crystals (Toshiba Ceramics, Inc.) and placed onto a $SiO_2$ on Si substrate as described in ref. [1]. Graphene sheets with lateral sizes as large as 50 μm were placed in contact with Au grounding stripes via thermal deposition through a metal shadow-mask; the shadow-mask technique avoids contamination of the graphene sheets with photoresist. The graphene layers were characterized with a *microscopic* approach, combining methods sensitive to surface structure—micro-spot low energy electron diffraction (μLEED) and low energy electron microscopy (LEEM)—with micro-spot angle-resolved photoemission spectroscopy (μARPES) to probe the electronic structure. The experiments were performed in UHV conditions at the Nanospectroscopy beamline in use at the Elettra Synchrotron facility in Trieste, Italy. This instrument reaches a lateral resolution of less than 40 nm in imaging modes and has an energy resolution of 300 meV. μARPES and μLEED measurements were restricted to regions of 2 μm in diameter, which had previously been characterized by LEEM. Momentum resolution for ARPES experiments was 0.019 $Å^{-1}$. ARPES measurements were made with 90 eV photons. X-ray photoemission electron microscopy (XPEEM) with 403 eV photons was employed to examine the graphene sheets for impurities that may have been introduced in the preparation procedure. Traces of Au contamination were identified only in the immediate proximity (< 5 μm) of the Au stripes (see Fig. 1). The LEED and ARPES experiments described below were restricted to areas of graphene that were not contaminated by Au deposition.

LEEM was used to locate sample areas of interest and determine film thickness. Figure 1 compares optical microscopy and LEEM images of one of the samples used in the experiments described below. The grey bands in the LEEM image correspond to graphene regions of different layer thickness. The observed difference in contrast is a quantum-size effect resulting from the interference between electron waves scattered at the surface and at the interface with the substrate. The positions and number of the maxima and minima in the electron reflectivity as a function of electron energy allow the identification of the exact film thickness [10,11,12,13] (details will be presented elsewhere). The sample-layer thickness was independently confirmed by micro-Raman measurements [14,15,16].

The crystalline structure of the samples was investigated using μLEED. A careful analysis of LEED intensity as a function of electron energy (details will be presented elsewhere) indicates that multilayer graphene and graphite samples have the 3-fold symmetry expected for A-B-A stacked graphitic solids, while the monolayer LEED pattern displayed the expected 6-fold symmetry. As shown in Fig. 2, the Gaussian spread of the primary and secondary LEED peaks increases with decreasing film thickness, reaching a maximum for monolayer graphene. Additionally, we note that the Gaussian width of the diffraction spots increases linearly with $k$, where $k$ is the total momentum of LEED electrons, related to LEED energy by $k = \sqrt{2m_e E_{kin}}$, which suggests that the surface of the graphene layers is microscopically corrugated. If we make the simplifying assumptions that the corrugation of graphene is random and that the length scale of the ripples is larger than the transfer width of our LEED apparatus, we may model our data as an incoherent sum of LEED intensities over multiple domains with different local surface normals. Assuming a Gaussian distribution for the local surface normal, the standard deviation, $\Delta\theta_{norm}$, can be readily obtained using a simple trigonometric relation:

$$\Delta\theta_{norm} \approx \frac{1}{2}\frac{\hbar\Delta k_{||}}{\sqrt{2m_e E_{kin}}} \quad (1)$$

where $\Delta k_{||}$ is the Gaussian width of the diffracted beam. As shown in the table below, our measurements put an upper bound on the variation in surface normal of about 6$^{o}$ for monolayer graphene, which is in accord with recent results using scanning tunneling microscopy and electron diffraction [17,18,19,20].

| Thickness (ML) | $\Delta k_{\parallel}$ (Å$^{-1}$) | $\Delta\theta_{norm}$ (deg) |
|---|---|---|
| 1 | 0.70 | 6.1 |
| 2 | 0.28 | 2.4 |
| 3 | 0.20 | 1.7 |

$\Delta\theta_{norm}$ as a function of layer thickness calculated using eq. (1)

In the standard tight-binding model, the theoretical dispersion of graphene valence band electrons is given by [21]:

$$E(\mathbf{k}) = -t\sqrt{1+4cos\left(\sqrt{3}ak_y/2\right)cos\left(ak_x/2\right)+4cos^2\left(ak_x/2\right)} \quad (2)$$

where $t$ is the nearest neighbor hopping energy and $a$ is the lattice constant. This equation yields a nearly linear dispersion relation in the vicinity of $E_F$—electrons behave like massless particles, traveling at a fixed speed of ~10$^6$ m/s. Using Angle-resolved photoemission spectroscopy (ARPES) we are able to directly probe this dispersion. We collected ARPES data from graphite flakes and monolayer graphene sheets over the whole Brillouin zone from 1 eV to –20 eV (referenced to $E_F$). Comparing the angle-integrated graphene and graphite spectra in Fig. 3 (c) we note that although the overall features are similar, the graphene spectrum is shifted towards higher binding energy by ~300 meV, an effect which we attribute to charging of the $SiO_2$ substrate. The angle-resolved graphite spectrum is in accord with prior measurements of the graphite band structure [22,23]. The degraded appearance of the angle-resolved monolayer spectrum is due, first, to the undulations of the graphene layer, which produce broader photoemission features than the atomically flat graphite flakes. Additionally, the 3.4 Å thick graphene film is semi-transparent to UV photons and emitted photoelectrons at these energies [24]. Thus, our measured photoemission spectrum contains contributions from $SiO_2$ photoelectrons. Specifically, a non-dispersive $SiO_2$ peak appears in the single layer spectrum at –7 eV, which is in accord with previous measurements of the amorphous $SiO_2$ band structure [25,26]. Although $SiO_2$ photoemission obscures features of the graphene spectrum below this peak, the large 9 eV $SiO_2$ band gap [25] allows us to study the graphene valence band in the vicinity of $E_F$.

Momentum distribution curves (MDC) in the vicinity of the K point for both graphite and graphene are shown in Fig. 4. Due to an interference effect [23], only one dispersing branch is seen in the ΓK direction, while two symmetric branches appear in the ΓM direction. The graphite MDC is well fit by two Lorentzian curves representing the split π band associated with the two inequivalent graphite sub-lattices. By contrast, the graphene MDCs are well fit by a single peak since the two sublattices in graphene are degenerate. Note that the graphene MDCs are wider than the graphite MDC by a factor of ~4. In ARPES experiments, MDC width can normally be related to the complex self energy, $\Sigma(\mathbf{k})$, of charge carriers in the crystal [27,28,29]. However, in our experiment there are contributions to MDC width that are unrelated to $\Sigma(\mathbf{k})$. The first is interaction with charged impurities in the $SiO_2$ substrate, which results in a local shift in the chemical potential of the graphene sheet—an excess of electrons or holes often called a "charge puddle" [30,31]. Since the lateral scale of these puddles is ~100 nm [31], our photoemission results (collected over ~2 μm) average over many such fluctuations, causing a broadening in energy space of our photoemission features. The corresponding broadening in k space is then given by $\Delta k_{\|} = \Delta E / \eta v_F$ (assuming linear dispersion). $\Delta E$ is estimated to be ~100 eV [30,31]; from this we obtain, $\Delta k_{\|}$ = 0.015 Å$^{-1}$, which is a small contribution to the total MDC width of 0.47 Å$^{-1}$.

Undulations on the graphene surface, on the other hand, are expected to produce a larger contribution to MDC broadening. As with our LEED results, our ARPES measurements sample many tilted domains since the characteristic length scale for the ripples is two orders of magnitude smaller (~10 nm) [19] than our photoemission region (~2μm). If we model our ARPES results as an incoherent sum over multiple domains we may again use equation (1) to estimate $\Delta\theta_{norm}$ from the spread in $k_{\|}$. For electrons photoemitted from the Fermi level $E_{kin} \approx 86$ eV and the fits in Fig. 4 provide $\Delta k_{\|}$ = 0.47±0.03 Å$^{-1}$ which yields $\Delta\theta_{norm}$ = 3°. It is not surprising that this value for $\Delta\theta_{norm}$ is smaller than that calculated from the LEED data, since the electron phase coherence length in graphene is expected to be longer than the length scale for surface undulations. Thus, the domains may not be treated independently. A more sophisticated theoretical approach is necessary to understand the effect of surface undulations on the photoemission data.

Despite the broadening of photoemission features described above, it is possible to accurately measure the electronic dispersion near the K point. The graphene MDCs at regular energy intervals were fit with a single Gaussian peak in the ΓK direction and a sum of two Gaussian peaks in the ΓM direction. The single peak Gaussian width obtained from the fits is $0.47 \pm 0.03$ Å$^{-1}$ in both parallel momentum directions, independent of energy. MDC peak height has been normalized in Figs. 4(b), (e) and (f) to emphasize the dispersion. However, the peak height was observed to decrease linearly with increasing energy, reaching a minimum for MDCs taken near the Dirac point, indicating that the microcleaving process yields high-quality crystalline graphene sheets, which are free from defects. Peak positions obtained from the fits are indicated in the Figs. 4(b) and 4(c) along with the theoretical band dispersion predicted by the one-orbital tight binding model. The Fermi velocity extracted from the best fit lines is $1.0 \pm 0.15 \times 10^6$ m/s, in excellent agreement with the theoretical prediction.

A close look at the peak positions shown in Fig. 4(c) reveals that the Dirac point is ~250 meV below the Fermi level. This may be attributed to charging of the $SiO_2$ substrate by photoionization from the incident UV beam. As described above, the graphene layer is sufficiently thin to allow incident UV photons to reach the substrate and eject photoelectrons, leaving pockets of trapped positive charge, which results in an overall shift of the chemical potential. As noted above, the same effect can be observed by comparing the angle-integrated graphene and graphite spectra shown in Fig. 3(c), which yields a similar result for the shift.

In summary, we have used spectro-microscopy to probe the surface morphology and electronic structure of single and multilayer graphene on a micron scale. Our LEED results show that the characteristic graphite hexagonal lattice structure is maintained up to the single layer. However, electron diffraction is degraded by undulations of the graphene surface. An estimate of the standard deviation in surface normal can be obtained from an analysis of the LEED data—our result is comparable to recent measurements of the same quantity by STM and TEM. Photoemission measurement of the valence band of monolayer graphene is possible due to the large bandgap of the $SiO_2$ substrate. The MDC widths of graphene peaks are larger than those of comparable features on graphite, an effect that is attributed to undulations of the graphene layer. Nonetheless, an accurate measurement of the electronic dispersion near $E_F$ is possible. Our data show clearly that the band structure of exfoliated graphene is well described by the standard one-orbital tight binding model, which predicts a linear dispersion relationship near the K point. Our experimentally determined Fermi velocity is equal to the theoretical prediction of $10^6$ m/s within a small margin of error of ~15%. These results are of paramount importance for future work in the field as this is experimental verification of massless fermionic dispersion in exfoliated graphene. A greater theoretical understanding will be necessary to extract the complex self energy, $\Sigma(\mathbf{k})$, from the MDC widths as undulations of the graphene layer obscure this quantity. Additionally, further experimental work will be necessary to understand the effect of the $SiO_2$ substrate on these undulations.

This research (RO) was supported by the DOE (Contract No. DE-FG02-04-ER-46157) and by (KK, PK, SW) NSF Award Number CHE-0641523 and by NYSTAR. The synchrotron portion of the project (AM, DC ,AL, TM, MN) was supported through MiUR-PRIN2006-prot. 2006020543_002. AM gratefully acknowledges the NSEC and the Italian Academy at Columbia University for the warm hospitality and financial support during his visit.

---

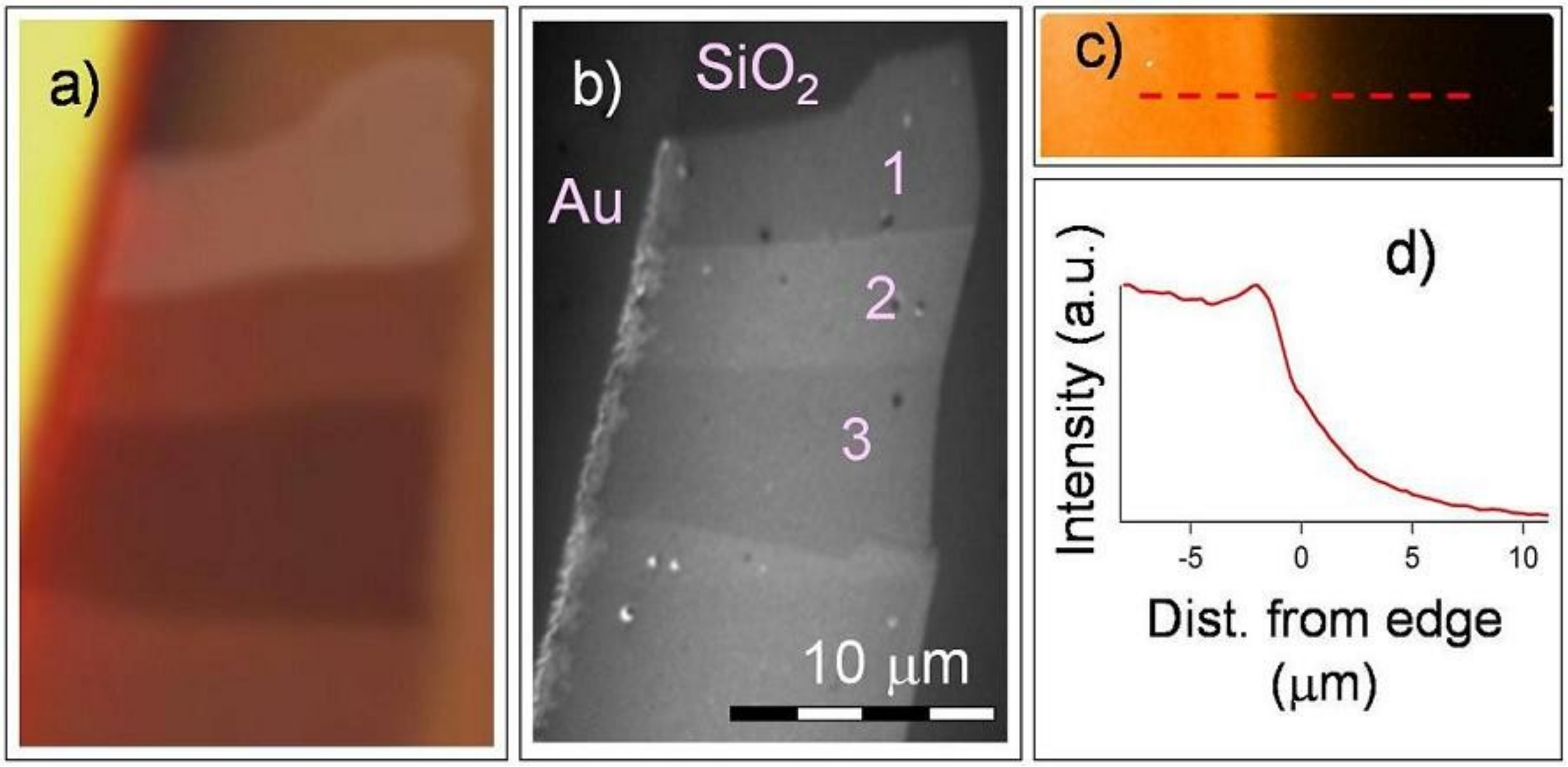

FIG. 1 (a) Optical microscopy image of monolayer and multilayer graphene sample. (b) LEEM image of the same sample (numbers indicate graphene sample thickness in ML). (c) XPEEM image of Au 4f 7/2 core level taken at edge of Au wire on graphene. (d) Intensity profile along direction indicated by dashed line in (c).

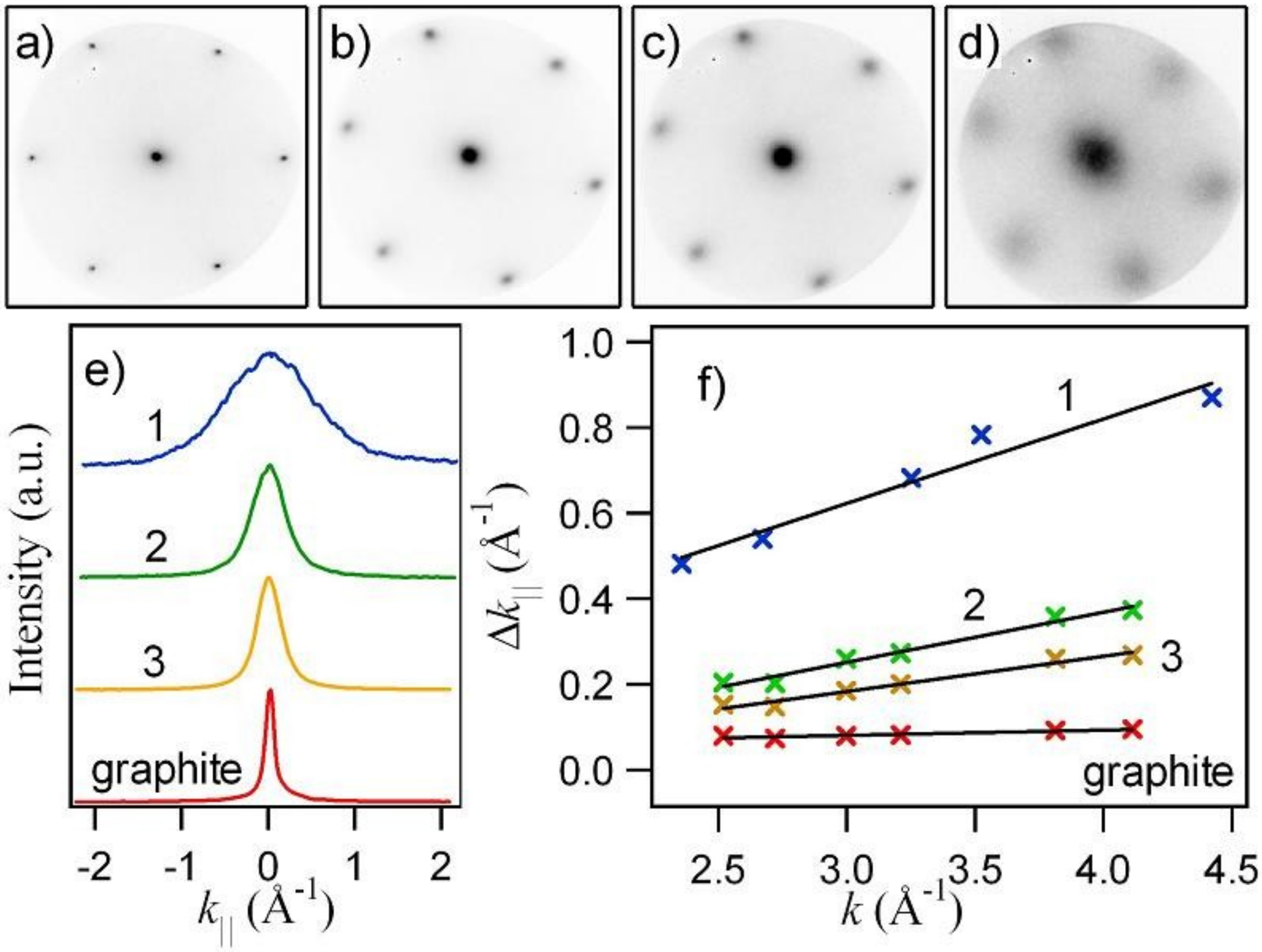

FIG. 2 (a-d) Graphite, trilayer, bilayer, and monolayer graphene LEED patterns at 42 eV, respectively. (e) Intensity profiles of central diffraction maximum from (a-d). (f) Gaussian width of central diffraction maxima as a function of *k*.

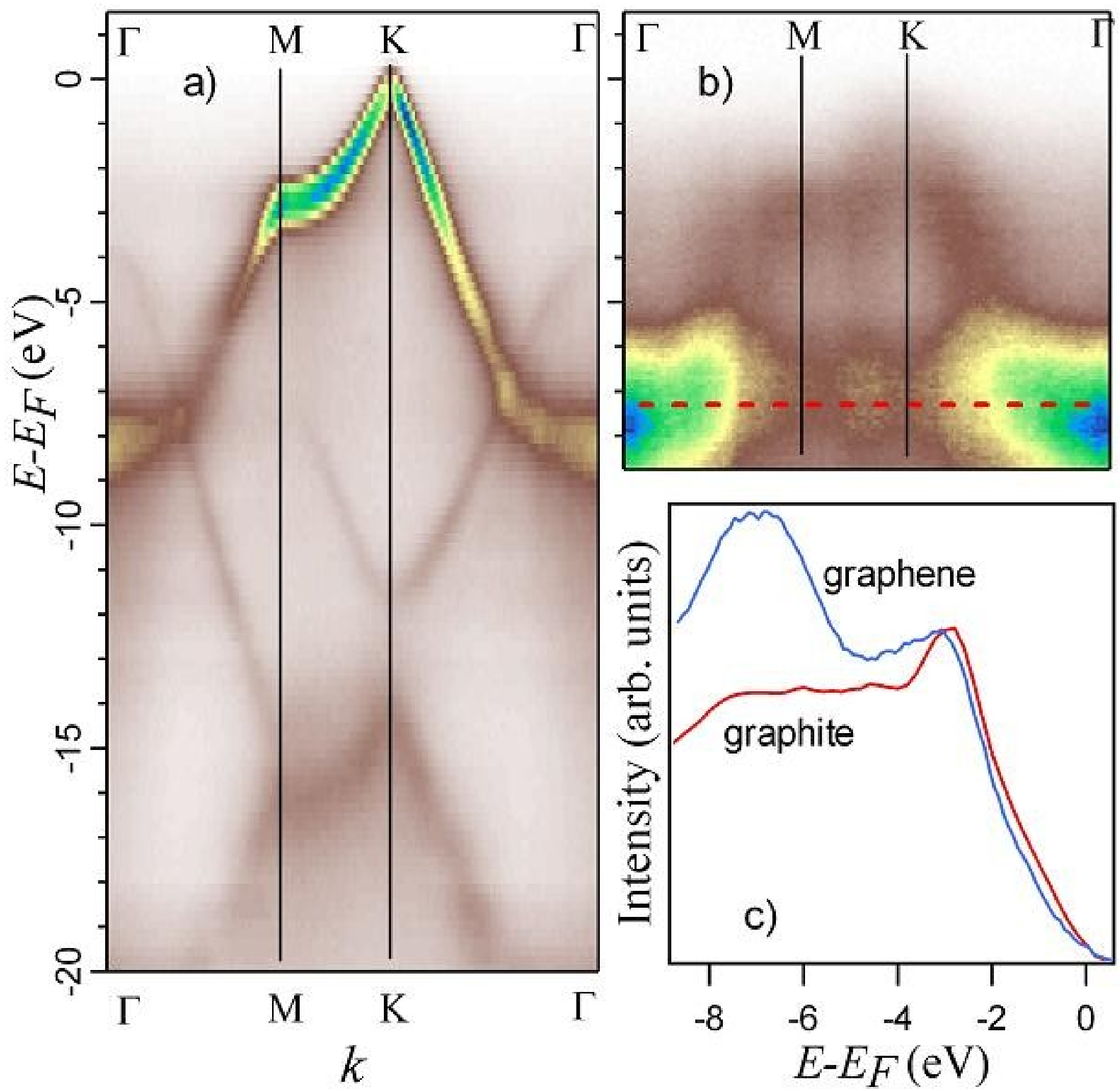

FIG. 3 (a-b) ARPES intensity along principal Brillouin zone directions for graphite and graphene, respectively (Brillouin zone symmetry-point labels are valid for a 2D approximation of the graphite Brillouin zone). (c) Angle integrated photoemission intensity.

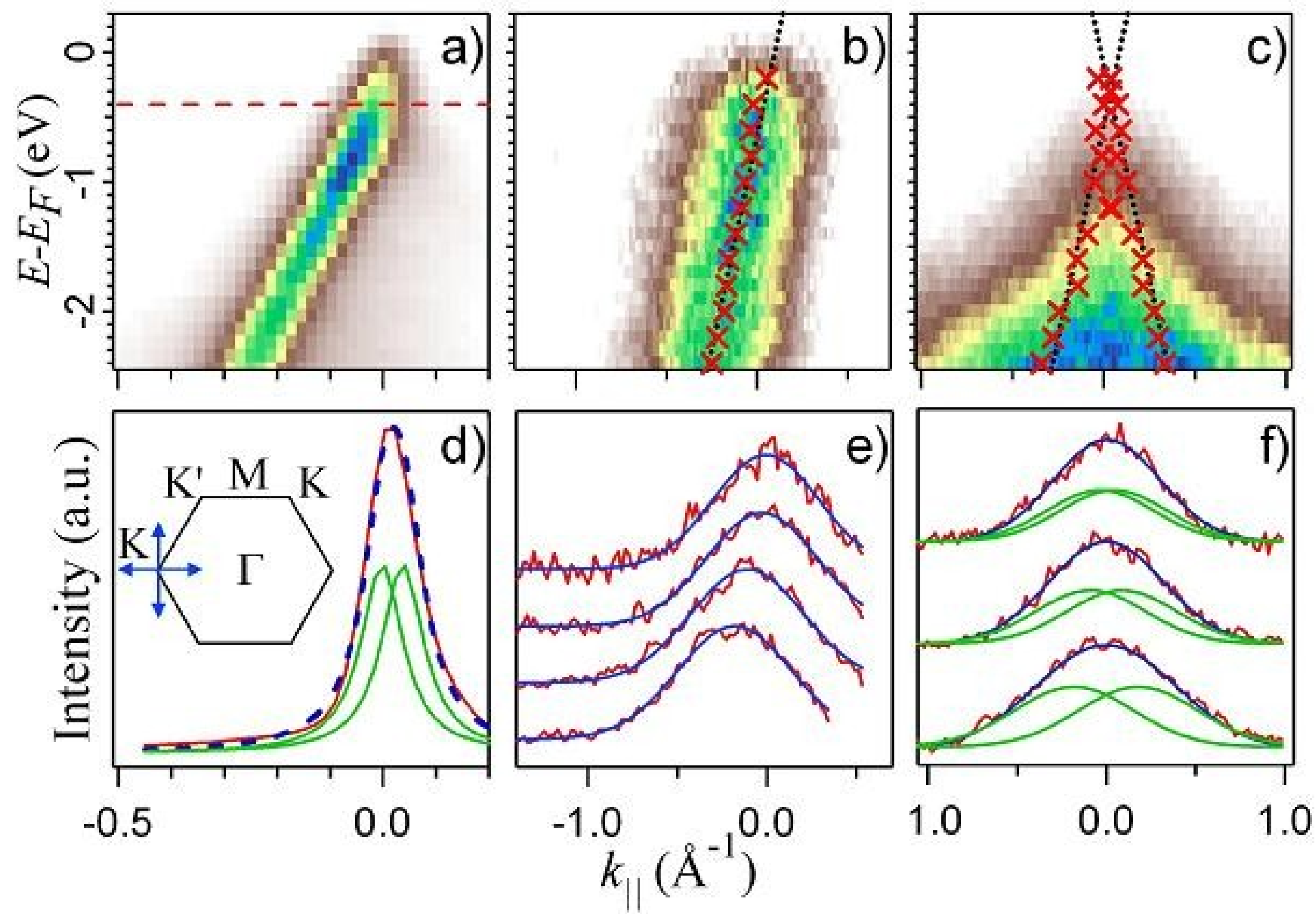

FIG. 4 (a-b) ARPES intensity through K point along ΓK direction for graphite and graphene, respectively (momentum referenced to K point). (c) Graphene ARPES intensity through K point along ΓM direction. (d) MDC along dotted line in (a) (inset shows 2D graphite/graphene Brillouin zone, arrows indicate ΓK and ΓM directions). (e-f) MDC Gaussian peak fits to above graphs.